\documentclass[%
 reprint,
 amsmath,amssymb,
 aps,
]{revtex4-1}

\usepackage{graphicx}
\usepackage{dcolumn}
\usepackage{bm}

\usepackage{color}
\usepackage{natbib}
\usepackage[titletoc,toc,title]{appendix}
\usepackage{wrapfig}

\begin{document}

\preprint{APS/123-QED}

\title{Fate of False Vacuum in Superconducting Flux Qubits}

\author{Ali Izadi Rad}
 \altaffiliation[ ]{Physics Department, Sharif  University Of Technology.}
\author{Hesam Zandi}%
 \email{Zandi@ee.sharif.edu}
 \author{Mehdi Fardmanesh}
 \email{Fardmanesh@ee.sharif.edu}
\affiliation{%
 School of Electrical Engineering, Sharif University of Technology, Tehran, Iran
}%

\collaboration{Superconductor Electronics Research Laboratory (SERL)}

%
%

\begin{abstract}
We  propose  a similarity between the scenario of  fate of false vacuum in cosmology at early universe and  the situation in where the quantum state decays in superconducting flux qubits. This is due to the fact that  both cases have two homogeneous stable equilibrium states in scalar field, which in quantum theory, could penetrate through the barrier in different possibilities and hence considered unstable  decaying in time.
In quantum computation, decay rate is among the most important factors in characteristics of the system like coherency, reliability, measurement fidelity, etc.
In this considered potential, the decay rate from the penetrating (False vacuum) state to the stable (absolute minimum) state is achieved to leading order in $\hbar$ by the approach of  Instanton model.
In case of the superconducting flux qubit having thin barrier potentials, the decay rate is calculated and its relations with the actual set of parameters in flux qubit design are introduced.

\end{abstract}

\pacs{Valid PACS appear here}
\maketitle


\section{Introduction}
Many inflationary cosmological model of our universe contain some local minimum energies with unique absolute one. These models have included the possibility  of existing   a long-lived, but not completely stable, sector of space, which could potentially at some time transits to more stable vacuum state \cite{1001}\cite{1002}\cite{1003}\cite{1004}\cite{1005}\cite{1006}. 
Trace of similar idea exists in String theory and string landscape \cite{1011}\cite{1012}. The Standard model of particle physics discuses whether the universe's present electroweak vacuum state is likely to be stable or merely long-lived by considering the masses of the Higgs boson and the top quark,
\cite{1007}\cite{1008}.

The decay of the false vacuum was became serious by papers of Coleman and Callan \citep{1009},\citep{1010}. In the classical field theory there is a possibility to have two homogenous stable equilibrium states with different energy densities. However in the Quantum version, the tunneling effect and barrier penetration causes the one of the state become unstable and decays to the lower one. This unstable state called False vacuum.
%
Coleman shown that if the state of the early universe located in false vacuum and cannot transits  with fluctuations to the true vacuum state. But still the quantum tunneling can provide this ability for false vacuum to transits to true part of vacuum.
%
The unraveling idea of fate of false vacuum can be used in study of superconducting Phase Qubits. One of the promising candidates of building the quantum bits devices based on superconducting Josephson junction can be a promising candidate for future quantum computers\citep{1014},\citep{1015}. One category of Josephson Qubits is Fulx Qubits\citep{1016}. The connection of flux qubits with false vacuum  becomes clear when  we study the potential term in of Lagrangian  flux qubit.  The similarity of potential  fulx qubits  with  typical potential whihc presented in the scenario of false vacuum help us to calculated the decay rates of states in flux quibits. 
   We first review on physical structure of flux qubits, then  application of Feynman path integral in double well potential and estimation of  the ground state energy of the symmetric potential like one Josephson junction structure and rf-SQUID  has been introduced. In the next step, we use this approach   to estimate the decay rate of false vacuum in typical potentials. Finally we apply these calculations to flux qubits.

\section{Flux Qubit}

The RF-SQUID consists of the tunnel Josephson junction inserted in a superconducting loop, as illustrated in Fig.I This circuit realizes magnetic flux bias for the Josephson junction.
In the absence of dissipation, the Lagrangian of the RF-SQUID is given by
\begin{equation}
L(\phi,\dot{\phi})=\frac{{\hbar}^2 {\dot{\phi}}^2}{4E_C}-E_J(1-\cos{\phi})-E_L \frac{(\phi-{\phi}_e)^2}{2}
\end{equation}
Where ${\phi}_e=\frac{2e}{\hbar} \mathbf{\Phi}_e$ and $\mathbf{\Phi}_e$ is the external flux threading the SQUID loop.The last term in this equation corresponds to the energy of the current circulating in the loop
\begin{equation}
E_L=\frac{{\mathbf{{\Phi}_e}}}{4{\pi}^2 L}
\end{equation}

\begin{center}
\begin{figure}[h]\label{YaNabi007}
\includegraphics[height=35mm]{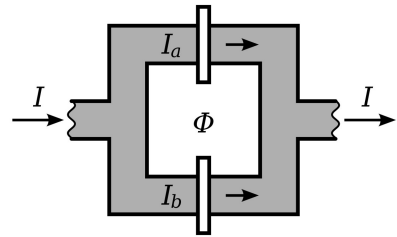}
\caption{
DC-SQUID, the suggested structure of for superconducting flux quibits
}
\end{figure}
\end{center}
%
%
%
The potential energy $U(\phi)$ corresponding to the last term in Eq.1 is schetched in Fig.II . 
\begin{center}
\begin{figure}[h]\label{YaNabi008}
\includegraphics[height=35mm]{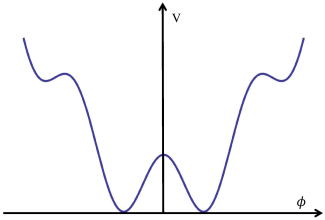}
\caption{
Flux potential in typical SQUIDs for one   sets of parameters
}
\end{figure}
\end{center}
Now if we put two Josephson junction in parallel to each other in the superconducting loop, compare to the single mode, the effective Josephson energy of double junction related to the magnetic flux passing from the SQUID loop. 
If we show the superconducting phase of two junction by ${\phi}_{1,2}$, we find that 
\begin{equation}
I_{c_1}\sin{\phi}_1-I_{c_2}\sin{\phi}_2=I_{e}
\end{equation}
By defining ${\phi}_1+{\phi}_2={\phi}_e$ and ${\phi}_{\pm}=\frac{{\phi}_1 \pm {\phi}_2}{2}$ it is easy to show that by this variables, the potential has the form 
\begin{equation}
U(\phi)=2E_J \cos(\frac{{\phi}_e}{2})(1-\cos{\phi}_{-})-\frac{\hbar}{2e}I_{e} {\phi}_{-}
\end{equation}
%
%
%
%
%
%
%
%

\section{Instanton Model for Symmetric Potentials}
In order to study the decay rate of the false vacuum in Flux superconducting Qubits, it is necessary to review on general analyze of one particle behavior in one-dimensional barrier potential using the method of path integral . If we consider the particle with unite mass which is under the influence of the one-dimensional potential V(x), then by following the Euclidean form of the Feynman path integral we can describe the evolution of the particle with 
\begin{equation}\label{9003}
\langle x_f | e^{-\frac{HT}{\hbar}}|x_i \rangle =N \int [dx] e^{-\frac{S}{\hbar}}
\end{equation} 
Here $|x_f \rangle $ and $|x_i\rangle $ are the eigenvalue of the space and the $N$ refer to normalization factor, $H$ represent the Hamiltonian of the system which can be depends on time and $T$ shows the time interval of the evolution.
the symbol of $[x]$ denotes the integration over the all functions $x(t)$ that obey from the boundary condition $x(-\frac{T}{2})=x_i, x(+\frac{T}{2})=x_f$ If $\bar{x}$ be any functions which obeys the boundary condition then we can write $x(t)=\bar{x}(t) +\sum_{n} c_n x_n(t)$ where the set $x_n$ build the complete set, $\int_{-\frac{T}{2}}^{\frac{T}{2}} x_n(t) x_m(t)={\delta}_{mn}$ and $x_n(\pm \frac{T}{2})=0$. By these condition we can rewrite the mesure of the integral by 
%
\begin{equation}
[dx]={\prod}_{n} (2\pi \hbar)^{-\frac{1}{2}}dc_n
\end{equation}
It is suitable to set $\bar{x}$ as the classical path and thus it must be the answer of variation of the action
\begin{equation}{\label{fate001}}
\frac{\delta S}{\delta \bar{x}}=-\frac{d^2 \bar{x}}{dt^2}+ V'(\bar{x})=0
\end{equation} 
If we select the complete set of $\{x\}$ as the solution of the following equation
\begin{equation}{\label{9001}}
\frac{d^2 x_n}{dt^2}+V''(\bar{x})x_n={\lambda}_{n} x_n 
\end{equation}
Therefore calculation in order of $\hbar$ and using the semi classical approximation lets us to write the evolution of the system by 
\begin{eqnarray}\label{equation001}
\langle x_f | e^{-\frac{HT}{\hbar}} |x_i \rangle &&=N e^{-\frac{S(\bar{x})}{\hbar}} {\prod}_{n} {{\lambda}_n}^{-\frac{1}{2}} [1+O(\hbar)] \\ \nonumber
&&= N e^{-\frac{S(\bar{x})}{\hbar}}[det(-{\partial}_{t}^2+V''(\bar{x})]^{-\frac{1}{2}}[1+O(\hbar)]
\end{eqnarray}
If we define ${\omega}^2$ to be $V''(0)$ then the standard calculation shows that for large $T$ 
\begin{equation}
N[det(-{\partial}_t^2+{\omega}^2)]^{-\frac{1}{2}}=(\frac{\omega}{\pi \hbar})^{\frac{1}{2}}e^{-\frac{\omega T}{2}}
\end{equation}

\begin{center}
\begin{figure}[h]\label{Elahi001}
  \includegraphics[height=35mm]{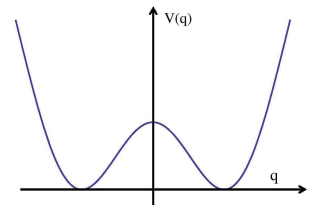}
  \caption{
 Double well potential which is symmetry respect to $q$, $V(q)=V(-q)$.
 }
  \end{figure}
  \end{center}

Now we discuss about the application of path integral to study the bounce situation, the motion particle in the presence of the potential like Fig \ref{Elahi001}. It is necessary to understanding the Instanton model for double well potential which more similar to Flux Qubits potential.


in the solid two approximately equal metastable minimum around the the ideal binding axis has been created and the energetically lowest excitations of the system are transitions of individual atoms between nearly degenerate minima and the atoms flips around the binding axis.
in the other hand there is well know shape of potential which is similar to double well potential and it is the Higg's potential with Lagrangian 
$ L=\frac{1}{2}{\partial}_{\mu}\phi {\partial}^{\mu} \phi -\frac{1}{2}m^2 {\phi}^2- g^2 {\phi}^4 $ and there was some efforts in 70th to describe some concept with tunneling via this potential.

 We assume that the potential is even 
\begin{equation}
V(x)=V(-x)
\end{equation}
And if we denote the minimum of potential with $a$ and $-a$ and the second derivation of potential at these points with $V''(\pm a)={\omega}^2$we would like to estimate the quantity like 
\begin{equation}
\langle -a|e^{-\frac{HT}{\hbar}}| -a \rangle =\langle a|e^{-\frac{HT}{\hbar}}| a \rangle
\end{equation}
Or 
\begin{equation}
\langle -a|e^{-\frac{HT}{\hbar}}| a \rangle =\langle a|e^{-\frac{HT}{\hbar}}| -a \rangle
\end{equation}

By approximating the functional integral by it's semi classical limit, we need to at first step to calculate the classical Euclidean equation of motion, consistent with our boundary conditions.more precisely, there are three different types of classical solutions that fulfill the condition to be at coordinates $\pm a$ at times 0 ot $T$ one of them is where the particle rest at on top of one of the hill, $a$ the second one corresponding the similar situation with $-a$. But this two trivial solutions which are that the particle stay on the top of the hills are not our interest we can see they add zero amount to classical action.however there is another potentially interesting solution, where the particle start it’s motion at the top of one hill at $t=-T/2$ and reach to the other top on time $t'=T/2$. Then we limit $T$ to infinity.

 The bounce motion's energy is  zero, thus the equation of motion takes the simple form 
\begin{equation}
\frac{dq}{dt}=(2V)^{\frac{1}{2}}
\end{equation} 
Or equivalently 
\begin{equation}
t-t_0=\int_{0}^{q} dq'(2V)^{-\frac{1}{2}}
\end{equation}
And $t_0$ is the constant of the integral. The $q$ verses the time albeit  depend on the exact shape of the potential but the while case have the general feature which near the the minimum points of potential we have 
\begin{equation}
a-q \sim e^{-\omega t}
\end{equation}
which as usual $V''(\pm a)={\omega}^2$. It seems they are well localized particles in scale of $1/{\omega}$ and the tunneling takes place on this time scale.
%



%
%

The solution which discused   is called ``instanton" . The idea is that these objects are very similar in their mathematic structure  what are called solition , particle-like solutions of classical filed theories but unlike the soliton they are structure in time and the "instant-" come from it. Polyakov suggest this structure at first \citep{19}-\citep{1}-\citep{12}-\citep{14}-\citep{18}.

Simply  by replacing the $t$ to $-t$ we can have anti-instanton. It is easy to obtain the classical action as we discussed before 
\begin{equation}
S_0=\int dt[\frac{1}{2}(\frac{dq}{dt})^{\frac{1}{2}}+V]=\int dt (\frac{dq}{dt})^2=\int_{-a}^{a} dq (2V)^{\frac{1}{2}}
\end{equation}
We can see here that the action for two trivial solution of classical path are zero because $\dot{q}=0$. Also notices that $S_0$ is determined by the functional profile of the potentiall $V$ and it doesn't depend on the structure of the classical solution.

The confinement of the insatnton configuration to a narrow interval of time has an important implication. This is the critical importance because it means that for large $T$, the instanton and anti-Instanton are not the only approximate solutions of the equation of motion, they are also approximate solutions consisting of strings of sidely separated instanton and anti-instanton. We should summing over all such configurations with n objects centered at $t_1,t_2,\cdots ,t_n$ where
\begin{equation}
-T/2<t_n< \cdots <t_2<t_1<T/2
\end{equation}

For $n$ widely separated objects $S$ is $nS_0$  to evaluating the determinant we consider it has the form 
\begin{equation}
(\frac{\omega}{\pi \hbar})^{\frac{1}{2}}e^{-\frac{\omega T}{2}}K^n
\end{equation}
we evaluate $K$ later \citep{1} . This ansaltz has the simple reason. If we consider the potential well which has the shape similar to even function like the quadratic form the we expect that the eigenvalue of energy be similar to harmonic oscillator eigenvalue which are $\hbar \omega$ . To approve it if we following the path integral formalism it is easy to verify that for the a bit general form of well potential $V$ with $V''(0)={\omega}^2$ we have 
\begin{equation}\label{sss}
\langle 0| e^{-\frac{HT}{\hbar}}|0\rangle =N[det(-{\partial}_t^2+{\omega}^2)]^{-\frac{1}{2}}[1+O(\hbar)]
\end{equation}
But we know from the harmonic oscillator propagator that 
\begin{equation}
N[det(-{\partial}_t^2+{\omega}^2)]^{-\frac{1}{2}}=(\frac{m \omega}{\pi \hbar})^{\frac{1}{2}}e^{-\frac{\omega T}{2}}
\end{equation}
Now, if we consider that the proper Hamiltonian has the Eigensate of energies $|n\rangle$ with eigenvalues $n$ thus 
\begin{equation}
H|n\rangle =E_n |n\rangle 
\end{equation} 

Then 
\begin{equation}
\langle q'|e^{-\frac{HT}{\hbar}}|q \rangle =\sum_{n} e^{-\frac{E_n T}{\hbar}}\langle q'|n\rangle \langle n | q \rangle 
\end{equation}

for large $T$ the leading term belong to the least Eigenvalue, $E_0$ and therefore $\langle q'|e^{-\frac{HT}{\hbar}}|q \rangle \sim e^{-\frac{E_0 T}{\hbar}} \langle q'|0\rangle \langle0  | q \rangle $ .comparing this result with Eq. \ref{sss} leads to 
\begin{equation}
E_0=\frac{1}{2}\hbar \omega [1+O(\hbar)]
\end{equation}
And 
\begin{equation}
|\langle x=0|n=0\rangle |^2=(\frac{\omega}{\pi \hbar})^{\frac{1}{2}}[1+O(\hbar)]
\end{equation}
These calculation approve our ansaltz.

Now in order to calculating the propagator 
 also we should notice we must integrate over the locations of the centers
\begin{equation}
\int_{-T/2}^{T/2}dt_1 \int_{-T/2}^{t_1}dt_2 \cdots \int_{T/2}^{t_n-1} dt_n =\frac{T^n}{n!}
\end{equation}

Also we are not free to distribute instantons and anti-instantons arbitrary. For example of we start out at $-a$ the first object we encounter must be an instanton the next one must be an anti-instanton and etc. If we add up all case we have 
\begin{equation}
\langle -a | e^{-\frac{HT}{\hbar}}|-a \rangle =(\frac{\omega}{\pi \hbar})^{\frac{1}{2}} e^{-\frac{\omega T}{2}} \sum_{even\, n } \frac{(K e^{-\frac{S_0}{\hbar}}T)^n}{n!}[1+O(\hbar)]
\end{equation}

while we have 
\begin{equation}
\langle \pm a | e^{-\frac{HT}{\hbar}}|-a \rangle =(\frac{\omega}{\pi \hbar})^{\frac{1}{2}} e^{-\frac{\omega T}{2}} \frac{1}{2}[e^{Ke^{-\frac{S_0}{\hbar}}T} \mp e^{-Ke^{-\frac{S_0}{\hbar}}T}]
\end{equation}

Therefore we see that we have two low-lying energy Eigenstates with energies 
\begin{equation}
E_{\pm}=\frac{1}{2}\hbar \omega \pm \hbar K e^{-\frac{S_0}{\hbar}}
\end{equation}

Where the $K$ coeficient has been explaned in Appendix.

\section{Fate Of False Vacuum  }

\begin{center}
\begin{figure}[h]\label{YaNabi002}
\includegraphics[height=45mm]{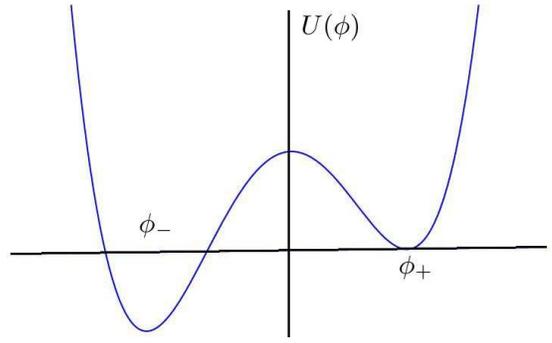}
\caption{
the typical potential for theory of fate of false vacuum
}
\end{figure}
\end{center}

The main and basic usage of the fate of false has been introduced in the single scalar field in four dimensional space time with no derivative interactions 
\begin{equation}
L=\frac{1}{2}{\partial}_{\mu} \phi {\partial}^{\mu} \phi -U(\phi)
\end{equation}
which the the potential possess two relative minima,${\phi}_{\pm}$, But only one of them, ${\phi}_{-}$ is the absolute minimum, Fig.\ref{YaNabi002}.
For this case the Euclidean equation of motion is 
\begin{equation}\label{9005}
[\frac{{\partial}^2}{{\partial \tau}^2}+{\nabla}^2]\bar{\phi}=U'(\bar{\phi})
\end{equation}
the boundary conditions for the bounce are $\lim_{\tau \pm \infty} \bar{\phi}(\tau,\mathbf{x})={\phi}_{+}$ and $\frac{\partial \phi}{\partial \tau}(0,\mathbf{x})=0$. Thus the Euclidean action given by 
\begin{equation}
B=\int d\tau d\mathbf{x} [\frac{1}{2}(\frac{\partial \phi}{\partial \tau})^2+\frac{1}{2}(\nabla \phi )^2+U]
\end{equation}
by the definition of new and useful variable 
\begin{equation}
\rho=[{\tau}^2+|\vec{\mathbf{x}}|^2]^{\frac{1}{2}}
\end{equation}
which Eq.\ref{9005} can be rewritten as 
\begin{equation}
\frac{d^2 \phi}{d{\rho}^2}+\frac{3}{\rho}\frac{d\phi}{d \rho}=U'(\phi)
\end{equation}
The action now can be written as
\begin{equation}
B=S_{E}=2{\pi}^2\int_{0}^{\infty} {\rho}^3 d\rho [\frac{1}{2}(\frac{d \phi}{d \rho})^2+U]
\end{equation}
This calculation in two-dimensional space time yields to 
\begin{equation}
B=S_{E}=2\pi \int_{0}^{\infty} \rho d\rho [\frac{1}{2}(\frac{d\phi}{d\rho})^2+U]
\end{equation}
\subsection{Decay Rate}
By looking up to Eq.\ref{equation001} we can see that this equation has the eigenvalue of zero and the corresponding Eigen function that provide this eigenvalue can be easily given by 
\begin{equation}
x_1=B^{-\frac{1}{2}}\frac{d\bar{x}}{dt}
\end{equation}
therefore we need to omit this troubling zero eigenvalue. we can use many ways to solve this problem and by the standard way we can calculate prime determinant which the zero eigenvalue has been omitted and we need to add coefficient to the $K$ like as $(\frac{B}{2\pi \hbar})^{\frac{1}{2}}$.
Now if we review on our solution way with more details we will find that as in one place the $\frac{d \bar{x}}{dt}$ become zero therefore $x$ has node and thus it cannot be the lowest eigenvalue of energy. It means that this system has the negative eigenvalue and the spectrum of our systems is the special case and we have unstable state here and the unitarily of this location form the Hilbert space can be in doubt.
By the valuable discussion in \citep{1009}\citep{1010} we find that we need to add the coefficient one-half to our calculation and the reliable result yields 
\begin{equation}
Im[N \int [dx]e^{-\frac{S}{\hbar}}]=\frac{1}{2}N e^{-\frac{B}{\hbar}}(\frac{B}{2\pi \hbar})^{\frac{1}{2}} T |\text{det}^{'} [-{\partial}_{t}^2+V''(\bar{x})]|^{-\frac{1}{2}}
\end{equation}
and by comparing the result with the definition of $K$ we will find that 
\begin{equation}
Im K =\frac{1}{2} (\frac{B}{2\pi \hbar})^{\frac{1}{2}} |\frac{det'[-{{\partial}_t}^2+V''(\bar{x})]}{det[-{{\partial}_t}^2+{\omega}^2]}|^{-\frac{1}{2}}
\end{equation}
n the stable situation, when the height of barrier penetration goes to infinity the solution of Schrödinger equation, corresponding to the ground state energy $E_0$ behaves as 
\begin{equation}
{\psi}_0(t) \sim e^{-\frac{iE_0t}{\hbar}}
\end{equation}
But for the case that we have not absolute minimum, $E_0$ becomes imaginary. Therefore for long times we have 
\begin{equation}
|{\psi}_0(t)|\sim e^{-\frac{Im E_0 t}{\hbar}}
\end{equation}
It clearly shows that the amplitude and therefore the probability of state decays. The parameter $|\frac{\hbar}{\text{Im}E_0}|$ is the lifetime of a now metastable state with wave function $\psi(t)$ . Let us to point out that the decay of state receives contributions from the continuation of all excited states. However, one expects, for intuitive reasons, that when the real part of the energy increases the corresponding contribution decreased faster with time, a property that can, indeed, be verified in examples. Thus, for large times, only the component corresponding to the pseudo-ground state survives. by considering the one-half calculation we have 
\begin{eqnarray}\label{9004}
\Gamma &&=-2 \text{Im}E_0/\hbar \\ 
&&=(\frac{B}{2\pi \hbar})^{\frac{1}{2}}e^{-\frac{B}{\hbar}} |\frac{det'[-{{\partial}_t}^2+V''(\bar{x})]}{det[-{{\partial}_t}^2+{\omega}^2]}|^{-\frac{1}{2}}[1+O(\hbar)] \nonumber 
\end{eqnarray}

\section{Fate of False Vacuum in Flux Qubits}
\begin{center}
\begin{figure}[h]\label{YaNabi010}
\includegraphics[height=55mm]{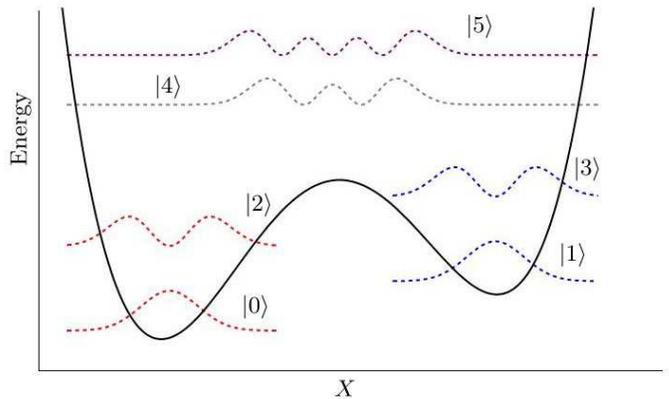}
\caption{
The states of Energy in superconducting flux qubits 
}
\end{figure}
\end{center}

In the Superconducting Flux Qubits as we discussed in the first section we have the potential with the form
\begin{equation}
U(\phi)=2E_J \cos(\frac{{\phi}_e}{2})(1-\cos{\phi}_{-})-\frac{\hbar}{2e}I_{e} {\phi}_{-}
\end{equation}
that has been sketched in Fig.\ref{YaNabi010}
We Can see that this potential is completely similar to potentials that we considered in the context of false vacuum. Now we are ready to applied results to this problem. the $U_{+}$ in this problem obviously is 
\begin{equation}
U_{+}(\phi)=2E_J \cos(\frac{{\phi}_e}{2})(1-\cos{\phi}_{-}) -\frac{\hbar}{2e}I_{e} a
\end{equation}
\subsection{The Thin Wall Approximation }
In the thin wall approximation we consider the original potential, $U_{+}(\phi)$, has the symmetric property
\begin{equation}
U_{+}(\phi)=U_{+}(-\phi)
\end{equation}
Which the potential has two absolute minima, $U_{+}(\pm a)=0$. Also it's is suitable to define $\mu$ as ${\mu}^2=U_{+}''(\pm a)$.
Now we break this symmetry by linear term as 
\begin{equation}
U=U_{+} \rightarrow U=U_{+}+\frac{\epsilon}{2a}(\phi -a)
\end{equation}
We Assume In the first order of $\epsilon$ the location of minimums are $\pm a$ again. This consideration is valid in these kind of potentials that we encounter in this paper. Here $\epsilon$ represent the differential value of potential in the two local minima. 
Here we can study the simple but important shape of potential which is similar to the Higgs's potential shape and it can be consider as firsts term of cosine function which we have in the Flux qubits's potential.
\begin{equation}\label{9006}
U_{+}=\frac{\lambda}{8}({\phi}^2-\frac{{\mu}^2}{\lambda})^2
\end{equation}
in this case by using the equation of motion we have
\begin{equation}
x=\int_{0}^{{\phi}_1} \frac{d\phi}{[2U_{+}(\phi)]^{\frac{1}{2}}}
\end{equation}
Thus the action will be 
\begin{equation}
S_{1}=\int dx [\frac{1}{2}(\frac{d{\phi}_1}{dx})^2+U_{+}]=\int_{-a}^{a} d\phi [2U(\phi)]^{\frac{1}{2}}
\end{equation}
Now by considering the $\mu|x|\gg 1$ we find that 
\begin{equation}
{\phi}_{1}=\pm(a-Ke^{-\mu |x|})
\end{equation}
this the general form of dependence of $\phi$ to $x$. but the analytical calculation for Mexicain hat potential, Eq.\ref{9006}, yields to ${\phi}_1=a \tanh(\frac{1}{2}\mu x)$ and the value of the action will be $S_{1}=\frac{{\mu}^3}{3\lambda}$.
\subsection{Estimating the Euclidean Action}
In terms of ${\phi}_{1}$, we can express analytically our approximation description of the bounce
\begin{eqnarray}
\phi &&= -a , \ \ \ \rho \ll R \\ \nonumber
&&={\phi}_{1}(\rho -R) , \ \ \ \rho \simeq R \\ \nonumber
&&=a , \ \ \ \rho \gg R \\ \nonumber
\end{eqnarray}
The only thing missing from this description is the value of $R$. we can find it by variation of the action respect the $R$.
Therefore it is easy to calculate the action by using the description of the $phi$ which yields to 
\begin{eqnarray}
S&&=2\pi \int_{0}^{\infty} \rho d\rho [\frac{1}{2}(\frac{d\phi}{d\rho})^2+U] \\ \nonumber
&&=-\pi R^2 \epsilon +2\pi R S_1
\end{eqnarray}
Variation respect to $R$ gives 
\begin{equation}
\frac{dS_E}{dR}=-2\pi R\epsilon +2\pi S_1=0
\end{equation}
Therefore 
\begin{equation}
R=\frac{S_1}{\epsilon}
\end{equation}
And finally we can obtain the Euclidean action 
\begin{equation}
B=S_E=\frac{\pi S_1^2}{\epsilon}
\end{equation}
Operationally this means that for desired shape of potential we need to calculate 
\begin{equation}
B=\frac{\pi}{\epsilon}[\int_{-a}^{a} d\phi [2U_{+}(\phi)]^{\frac{1}{2}}]^2
\end{equation}


Now, we can apply this calculation to our Flux qubit, according to potential of Flux qubit, the minima are located in 
\begin{equation}
\pm a=\pm \frac{\pi}{2}
\end{equation}
thus the $\epsilon $ will be 
\begin{equation}
\epsilon=\frac{\pi \hbar}{2e}I_e
\end{equation}
Now 
\begin{eqnarray}
x &&=\int_{0}^{{\phi}_1} \frac{d\phi}{[2U_{+}(\phi)]^{\frac{1}{2}}} 
\end{eqnarray}
Estimating the analytical description for ${\phi}$ is not straightforward and it is better to use the software to estimate it if we need it.
%
%
%

\section{conclusion}
In this paper, we discussed about the similarity of fate of false vacuum when we have two homogeneous stable equilibrium states in scalar field, which quantum version of field, made the one state unstable and it will decay to another one by barrier penetration. At first, we introduce the application of Feynman path integral in double well potential and estimate the ground state energy of the symmetric potential like one Josephson junction structure and rf-SQUID. this approach lets us calculate our desire parameters up to level $O(\hbar)$ of accuracy which called the semi-classical approximation. we achieve high gain estimate the decay rate of one state to next one, the states which have been selected as quantum levels in flux qubit. the scenario decay of false vacuum is completely similar to our problem. we applied this method to flux qubits's potential and we can do go more further if we use the thin well approximation. ultimately we show that the decay rate will be like 
\begin{equation}
\Gamma=f(B,U(\phi)) e^{-\frac{B}{\hbar}}[1+O(\hbar)]
\end{equation}
We declare how to estimate $f(B,U(\phi)) $ and $B$ for our superconducting flux qubits.

\begin{appendices}

\section{ Calculating the K Coefficient }
In order to evaluating the K coefficient, we should study the differential equation $[-d_{t}^2 +V''(x_c)]x=\lambda x$ with single instanton, we called the classical answer $\bar{x}$ . Because of time-translation invariance, this equation necessarily possesses an Eigen function of eigenvalue zero. It is easy to check that the following function satisfy this requirement 
\begin{equation}
x_1={S_0}^{-\frac{1}{2}}\frac{d\bar{x}}{dt}
\end{equation}
As it has the zero eigenvalue we should change the coordinate to omit this value. This action add Jacobian factor $\frac{S_0}{2\pi \hbar}$ to main formula. Now if we call the calculated determinant without zero eigenvalue by Prime notation then we have 
\begin{equation}
\langle a|e^{-\frac{HT}{\hbar}}|-a\rangle =NT(\frac{S_0}{2\pi \hbar})^{\frac{1}{2}}e^{-\frac{S_0}{\hbar}}(\text{det}[-{\partial}_{t}^2+V''(\bar{x})])^{-\frac{1}{2}}
\end{equation}
Therefor by comparing with what we defined we obtain 
\begin{equation}
K=(\frac{S_0}{2\pi \hbar})^{\frac{1}{2}}\sqrt{\frac{\text{det}(-{\partial}_{t}^2+{\omega}^2)}{\text{det'}(-{\partial}_{t}^2 +V''(\bar{x}))}}
\end{equation}
\section{Computing the deteriminant}
Here we want to study the differential equation likes the form of 
\begin{equation}
[-{\partial}_{t}^2+W]\psi=\lambda \psi
\end{equation}
Where $W$ is a bounded function of $t$ .we call ${\psi}_{\lambda}$ as the solution of our equation while boundary conditions are 
\begin{equation}
{\psi}_{\lambda}(-\frac{T}{2})=0 , \ \ \ \ \ \ \ \frac{\partial {\psi}}{\partial t}(-\frac{T}{2})=1
\end{equation}
The operator $[-{\partial}_{t}^2+W]$ has an Eigenvalue , $\lambda_{n}$, if and only if 
\begin{equation}
{\psi}_{\lambda_n}(T/2)=0
\end{equation}
Now, if $W^{(1)}$ and $W^{(2)}$ be two function of $t$ and the corresponding Eigen function of them be ${\psi}_{\lambda}^{1}$ and ${\psi}_{\lambda}^{2}$ the it is possible to proof that there is this the following statement \citep{1}
\begin{equation}
\text{det}[\frac{-{\partial}_{t}^2+W^{1}-\lambda}{-{\partial}_{t}^2+W^{2}-\lambda}]=\frac{{\psi}_{\lambda}^1(T/2)}{{\psi}_{\lambda}^2(T/2)}
\end{equation}
This valuable theorem help us to continue our complicated calculation. Know we want to estimate the determinant which the zero mode of it has been omitted, for this reason we estimate the determinant in the finite range of time $[T/2,-T/2]$ and divide it by its smallest eigenvalue, ${\lambda}_{0}$ and then letting to $T$ to go to infinity . Thus we should construct solution of $[-{\partial}_{t}^2+U''(x_c)]\psi=\lambda \psi$ where $x_c$ is the classical solution. We can estimate the solution of $\lambda=0$ , which is the classical solution. By defining the $S_0$ 
\begin{equation}
S_0=\int dt[\frac{1}{2}(\frac{dx}{dt})^2+U]=\int dt (\frac{dx}{dt})^2=\int dx\sqrt{2U }
\end{equation}
One of the solution is 
\begin{equation}
x_1=S_{0}^{-\frac{1}{2}}\frac{d x_c}{dt} \rightarrow Ae^{-|t|}, t\rightarrow \pm \infty
\end{equation}
Here the $A$ is constant. As we have second order differential equation then we should find another solution for it . we set the second solution which the value of Wronskian given by
\begin{equation}
x_1{\partial}_{t} y_1-y_1 {\partial}_t x_1=2A^2
\end{equation}
Which yields to 
\begin{equation}
y_1 \rightarrow \pm Ae^{|t|}\, , t \rightarrow \pm \infty
\end{equation}
To find the lowest eigenvalue we must find ${\psi}_{\lambda}(t) $ for small $\lambda$ .this can be done by standard method. We turn the differential equation to integral equation and iterate it once 
\begin{equation}
{\psi}_{\lambda}={\psi}_{0}(t)- {\lambda}(2A^2)^{-1} \int_{-T/2}^{t} dt'[y_1(t)x_1(t')-x_1(t)y_{1}(t')]{\psi}_{0}(t')
\end{equation}
Plus terms of the order ${\lambda}^2$ which we neglect. Thus 
\begin{equation}
{\psi}_{\lambda}(T/2)=1-{\lambda}(4A^2)^{-1} \int_{-T/2}^{T/2} dt[e^{T}x_{1}^2-e^{-T}y_{1}^2]
\end{equation}
For large $T$ the second term in integral can be neglected, thus 
\begin{equation}
{\psi}_{\lambda}(T/2)=1-\lambda (4A^2)^{-1}e^T
\end{equation}
Thus the lowest eigenvalue is given by 
\begin{equation}
{\lambda}_{0}=(4A^2)e^{-T}
\end{equation}
Therefor for large $T$ the modified determinant is given by 
\begin{equation}
\frac{det[-{\partial}_{t}^2+U''(x_c)]}{det[-{\partial}_{t}^2+{\omega}^2]}=\frac{{\psi}_{0}(T/2)}{\frac{{\lambda}_0e^{T}}{2}}=\frac{1}{2A^2}
\end{equation}

 \end{appendices}



\end{document}